\documentclass[twocolumn,lt]{jpsj3}
\usepackage{txfonts}
\usepackage{graphicx}
\usepackage{dcolumn}
\usepackage{bm}
\usepackage[usenames]{color} 
\usepackage{amssymb} 
\usepackage{amsmath} 
\usepackage[utf8]{inputenc} 
\usepackage{braket}
\usepackage{mathptmx}
\title{Anomalous Spin Transport Properties of Gapped Dirac Electrons with Tilting}

\author{
\name{Masao \surname{Ogata}$^{1,2}$}, 
Soshun \surname{Ozaki}$^1$, and Hiroyasu \surname{Matsuura}$^1$}
\inst{
$^1$Department of Physics, University of Tokyo, Hongo, Bunkyo-ku, Tokyo 113-0033, Japan 
\\
$^2$Trans-scale Quantum Science Institute, University of Tokyo, Hongo, Bunkyo-ku, Tokyo 113-0033, Japan
} 

\abst{
The anomalous spin transport coefficients of gapped Dirac electrons are studied with application to
a quasi--two--dimensional organic conductor $\alpha$-(BETS)$_2$I$_3$ in mind. 
In the presence of a gap induced by spin--orbit interaction, 
we show that the effective Hamiltonian is 
similar to the model considered by Kane and Mele with additional tilting. 
With this effective Hamiltonian, conductivity tensors up to the linear order of 
the applied magnetic field are obtained analytically using the microscopic linear response 
theory or Kubo formula. 
It is shown that spin Hall conductivity and anomalous diagonal spin conductivity 
proportional to the magnetic field become nonzero in this system, which are written in terms 
of the Berry curvature and orbital magnetic moment. 
The estimated values of spin conductivities using typical parameters turn out to be comparable to
the spin Hall conductivity in Pt.
}

\begin{document}
\maketitle

%
Massless and massive Dirac electron systems have peculiar physical 
properties such as orbital magnetism and transport. 
One of the most prominent phenomena is the large orbital magnetic susceptibility for 
bismuth,\cite{BiExp4,FukuyamaKubo,BiReview} 
three-dimensional Dirac systems,\cite{Suetsugu}
graphene,\cite{McClure,KoshinoAndo1,KoshinoAndoSSC} 
and related organic conductors.\cite{Fukuyama,KobayashiFuku}
Another prominent property is transport such as diagonal conductivity 
$\sigma_{xx}$\cite{ShonAndo,Gorbar}
and Hall conductivity.\cite{Gusynin,Fukuyama,FuseyaPRL}

More recently, topological contributions such as the Berry curvature have been discovered 
first in the transport properties\cite{TKNN,Kohmoto,Haldane,KaneMele,Niu,NiuRev,%
3D1,3D2,Konye}
and then in the magnetic susceptibility\cite{Xiao,Shi,Gao,Piechon,OgataTOD,Nomura,Ozaki} 
and have attracted considerable attention. 
In the transport properties, the relation with the Berry curvature has been discussed 
in the quantum Hall effect\cite{TKNN,Kohmoto,Haldane,Niu,NiuRev} and 
the quantized spin Hall effect in the gapped graphene model.\cite{KaneMele}
Recently, the effects of the Berry curvature on the conductivity, $\sigma_{\mu\nu}$ 
($\mu, \nu=x,y,z$), in the linear order of the applied magnetic field 
(including the Hall conductivity) has been clarified.\cite{3D1,3D2,Konye}

In this work, we study such transport properties in a two--dimensional massless and massive 
Dirac electron system with tilting, having the quasi--two--dimensional 
organic conductors 
$\alpha$-(BEDT-TTF)$_2$I$_3$\cite{OrganicRev,Kajita,Tajima,Katayama,Kino,Kobayashi,Hirata,Hirata2,Winter,Osada} and 
$\alpha$-(BETS)$_2$I$_3$\cite{Kato,Inokuchi,Hiraki,Kitou,TsumuSuzu,SuzuTsumu,Fujiyama} in mind. 
Since sulfur in the BEDT-TTF molecule is replaced by selenium in BETS, 
the latter has a stronger spin--orbit interaction than the former.\cite{Winter}
It has been established that these compounds have Dirac electron systems.\cite{OrganicRev}
The merits of studying these organic compounds are as follows. 
(1) Since they are quasi--two--dimensional, the bulk physical quantities 
can be accessible experimentally. 
(2) It is rather easy to tune the physical parameters such as the 
transfer integrals of electrons or the band structure\cite{KinoFuku,Seo,Hotta}
by applying pressure or by replacing the molecules constituting these materials. 
(3) The Dirac point, at which two linear band dispersions touch with each other 
in the massless case, is located very close to the Fermi energy because of 
stoichiometry, and the density of doped electrons or holes 
is considered to be very small.\cite{KobayashiFuku} 
(4) The two Dirac cones tilt in opposite directions, which gives an additional 
degree of freedom, sometimes leading to anomalous transport properties. 

However, in the massless Dirac electron systems in two dimensions, 
the Berry curvature vanishes.
In contrast, it does not vanish when a gap opens at the Dirac point 
owing to inversion symmetry breaking\cite{Suzumura,SuzuKoba,Hasegawa,Matsuno,Kitayama} 
or spin--orbit coupling.\cite{KaneMele,Winter,Osada,Kitou,TsumuSuzu}
In $\alpha$-(BEDT-TTF)$_2$I$_3$, a gap due to the charge ordering appears, which breaks 
the inversion symmetry. 
In this case, however, the Berry curvature near one of the two Dirac points, e.g.,
$\Omega_{{\bm k}_0}$, exactly cancels with $\Omega_{-{\bm k}_0}$ 
near the other Dirac point (i.e., $\Omega_{-{\bm k}_0}=-\Omega_{{\bm k}_0}$).
Thus, only the valley Hall effect is possible.\cite{Matsuno}
If we create an imbalance between the electron occupations near the two Dirac points 
by applying an electric 
field, this cancellation is broken and a finite contribution of the Berry curvature 
appearing in the nonlinear response theory\cite{Fu,OsadaD} 
will be detected.
On the other hand, when the gap opens as a result of the spin--orbit interaction, as in 
$\alpha$-(BETS)$_2$I$_3$,\cite{Kitou,TsumuSuzu} the above 
cancellation does not occur since $\Omega_{-{\bm k}_0}=\Omega_{{\bm k}_0}$.
However, we must take into account the spin dependence, which we discuss below.

We consider the following two--dimensional effective Hamiltonian 
for the Dirac electrons at a Dirac point, ${\bm k}_0$,
\begin{equation}
{\mathcal H} = -v\hbar t k_x + v\hbar (k_x \tau_x + k_y \tau_y) + \sigma_z \Delta \tau_z.
\label{Hamiltonian}
\end{equation}
Here, $\bm k =(k_x, k_y)$ is the momentum measured from ${\bm k}_0$. 
The first term represents the tilting with a parameter $t (0\le t \le 1)$, $v$ is the velocity of 
Dirac electrons, $\tau_i$ ($i=x,y,z$) are the Pauli matrices representing 
two sublattices, $\sigma_z$ is the $z$-component of spin, 
and $\sigma_z\Delta$ represents the gap induced by the spin--orbit interaction 
as considered by Kane and Mele in a graphene-type model.\cite{KaneMele}
We take $\Delta>0$. 

First, we discuss the origin of the $\sigma_z$ dependence of the gap $\sigma_z \Delta$ 
in the case of quasi--two--dimensional systems. 
In general, the spin--orbit Hamiltonian is given by
\begin{equation}
{\mathcal H}_{\rm spin-orbit} 
=\int d{\bm r} \psi^\dagger({\bm r}) 
\frac{\hbar}{4m^2c^2} {\bm \sigma}\cdot \left( {\bm \nabla} V({\bm r}) \times 
{\bm p} \right) \psi({\bm r}),
\label{SOC}
\end{equation}
which is derived from the relativistic Dirac equation. 
Here, $m$ is the bare electron mass, $c$ is the light velocity, 
$V({\bm r})$ is a periodic potential, and $\bm p$ is the momentum operator. 
In the quasi--two--dimensional systems, $V({\bm r})$ is approximately an even 
function with respect to $z$, and the dominant terms of the electron momentum 
$\bm p$ will be in the $x$-$y$ plane. 
Therefore, we expect that the integral of eq.~(\ref{SOC}) is small when it contains 
$\partial V/\partial z$ or $p_z$, which means that the dominant contribution of 
eq.~(\ref{SOC}) 
is proportional to $\sigma_z$.\cite{Ozaki,Osada}

The energy dispersion of eq.~(\ref{Hamiltonian}) is given by 
$E_\pm=-v\hbar t k_x \pm E_{\bm k}$ with 
$E_{\bm k}= \sqrt{\Delta^2 + v^2\hbar^2 k^2}$, 
and the Berry curvature is 
\begin{equation}
\Omega_{xy}^\pm =\mp \frac{1}{2} \frac{{\bm h}\cdot (\partial_x {\bm h}
\times \partial_y {\bm h})}{h^3} = \mp v^2 \hbar^2 \frac{\sigma_z \Delta}{2E_{\bm k}^3}, 
\label{BerryM}
\end{equation}
where $\partial_{x,y}=\partial/\partial k_{x,y}$ and ${\bm h}=(h_x, h_y, h_z)$ 
are the coefficients of $\tau_i$ in the Hamiltonian. 
In the present case, $(h_x, h_y, h_z)=(v\hbar k_x, v\hbar k_y, \sigma_z \Delta)$. 
%
Since we assume that the system has time-reversal symmetry, the 
effective Hamiltonian $\mathcal H'$ at the other Dirac point (i.e., at $-{\bm k}_0$)
is obtained from eq.~(\ref{Hamiltonian}) by the time-reversal 
operation (${\bm k}\rightarrow -{\bm k}, {\mathcal H}\rightarrow {\mathcal H}^*$, 
and $\sigma_z\rightarrow -\sigma_z$), which leads to 
\begin{equation}
{\mathcal H}' = v\hbar t k_x - v\hbar (k_x \tau_x - k_y \tau_y) - \sigma_z \Delta \tau_z.
\label{Hamiltonian2}
\end{equation}
We can see that ${\mathcal H}'$ gives the same $\Omega_{xy}^\pm$ 
as $\mathcal H$.
Note that this situation is different from the case in which the gap opens 
as a result of the charge ordering discussed in 
$\alpha$-(BEDT-TTF)$_2$I$_3$.\cite{Suzumura,SuzuKoba,Hasegawa,Matsuno,Kitayama}


In the following, we calculate the conductivity tensor 
$\sigma_{\mu\nu}$ ($\mu, \nu=x,y$) 
for the Dirac electrons at a Dirac point, ${\bm k}_0$, (i.e., per valley)
up to the linear order of the applied magnetic field $B$ using the Kubo formula. 
The current operators in the Hamiltonian (\ref{Hamiltonian}) are given by
$j_x = ev(-t+\tau_x)$ and $j_y = ev \tau_y$ ($e<0$). 
We consider the case with simple random impurities that have delta-function potentials. 
In this case, the vertex corrections due to impurity scattering are safely neglected,  
and the relaxation rate appearing as a self-energy in the Green's functions can be 
approximated as a constant $\Gamma$.
Note that, in the case of gapless Dirac electrons (or in graphene), 
Shon and Ando\cite{ShonAndo} 
showed that the energy dependences of the self-energy are important because of 
the $|\varepsilon|$ dependence of the density of states. 
However, when there is a gap as in the present case, we consider that the constant 
$\Gamma$ will be a reasonable approximation. 

First, in the absence of a magnetic field, we obtain the conductivity as
\begin{equation}\begin{split}
\sigma_{\mu\nu}^{(0)} 
&=-\frac{e^2}{\hbar L^2} \sum_{{\bm k},\sigma_z,\pm} 
\biggl[ \frac{f'(E_\pm)}{2\Gamma} \partial_\mu E_\pm  \partial_\nu E_\pm 
+f(E_\pm) \Omega_{\mu\nu}^\pm + O(\Gamma^1) \biggr],
\label{eq:sigma0B}
\end{split}\end{equation}
where $L^2$ is the two--dimensional area, 
$\mu, \nu=x$ or $y$, $f(\varepsilon)$ is the Fermi distribution function,
$f(\varepsilon) = 1/(e^{\beta(\varepsilon-\mu)}+1)$, 
and we have made an expansion with respect to $1/\Gamma$. 
[Although the derivation is equivalent to that by Karplus and Luttinger\cite{KarplusLuttinger} 
or by Thouless et al.,\cite{TKNN}  
it is shown in the Supplemental Material\cite{SM} for completeness.]
The first term of eq.~(\ref{eq:sigma0B}) gives the Drude-type conductivity.
The second term in the case of $\mu=x, \nu=y$ is the contribution first obtained by 
Karplus--Luttinger in the discussion of the anomalous Hall effect\cite{KarplusLuttinger}, 
although the concept of the Berry curvature was not known at that time. 
In the present Hamiltonian (\ref{Hamiltonian}), 
$\sigma_{xy}^{(0)}$ has only the contribution from the second term. 
However, since $\Omega_{xy}^\pm$ is proportional to $\sigma_z$, $\sigma_{xy}^{(0)}$ 
vanishes owing to the spin summation. 

\begin{figure}
\includegraphics[scale=0.3]{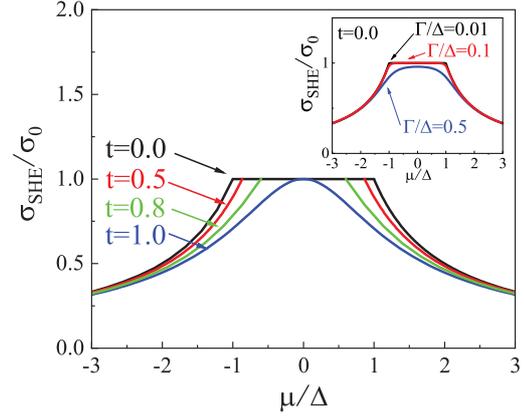}
\caption{(Color online) Chemical potential dependences of the spin Hall conductivity 
$\sigma_{\rm SHE}$ per valley at $T=0$ 
for some values of the tilting parameter $t$ ($t=0, 0.5, 0.8$ and $1.0$). 
The unit of $\sigma_{\rm SHE}$ is $\sigma_{0}=|e|/2\pi \hbar$. 
(Inset) $\Gamma$-dependence for the case $t=0$.}
\label{f1}
\end{figure}

In contrast, the spin Hall conductivity 
$\sigma_{\rm SHE} = \sigma_{Sxy}^{(0)}$ will have 
a finite value. In this case, we calculate the spin current 
$j_{Sx} = \sum_{\sigma_z} v\sigma_z (-t +\tau_x)$ in the presence of the 
electric field in the $y$-direction in the linear-response theory. 
Since the spin $\sigma_z$ is a good quantum number in the present model, 
the result is simply given by
\begin{equation}\begin{split}
\sigma_{\rm SHE} 
&=\frac{|e|}{\hbar L^2} \sum_{{\bm k},\sigma_z,\pm} 
f(E_\pm) \sigma_z \Omega_{xy}^\pm + O(\Gamma^1) \cr
&=\frac{|e|v^2\hbar^2\Delta}{2\hbar L^2} \sum_{{\bm k},\sigma_z, \pm} 
(\mp) \frac{f(E_\pm)}{E_{\bm k}^3}  +O(\Gamma).
\label{eq:SHE0}
\end{split}\end{equation}
This is the spin Hall conductivity predicted by Kane and Mele for a gapped 
graphene\cite{KaneMele} but, in the present case, with a tilting.
At $T=0$, the summation over the two--dimensional $\bm k$ can be carried out analytically, 
which yields\cite{SM}
\begin{equation}\begin{split}
\sigma_{\rm SHE} 
&= \frac{|e|}{2\pi \hbar} 
\begin{cases}
\frac{\Delta}{\sqrt{\mu^2+t^2 \Delta^2}}+O(\Gamma), \qquad 
{\rm for } |\mu| \ge \sqrt{1-t^2}\Delta, \cr
1+O(\Gamma), \hskip 2truecm {\rm for } |\mu| < \sqrt{1-t^2}\Delta. 
\label{SHE1}
\end{cases}
\end{split}\end{equation}
Note that this is the spin Hall conductivity per valley and that 
the band minimum (maximum) of the conduction (valence) band is given by 
$\sqrt{1-t^2}\Delta$ ($-\sqrt{1-t^2}\Delta$) in the presence of tilting. 
Figure \ref{f1} shows the chemical potential dependences of $\sigma_{\rm SHE}$ 
for some values of the tilting parameter $t$. 
Its behavior is very similar to the case of bismuth, which has a three-dimensional gapped 
Dirac electron system.\cite{BiReview,FOF2,FOF3}
In the present case, if we multiply $\sigma_{\rm SHE}$ by $|e|$, it has a dimension 
of two--dimensional conductivity. 
When the chemical potential $\mu$ is inside the gap, its maximum  becomes a 
universal value $|e|\sigma_{\rm SHE}=e^2/2\pi\hbar$. 
To convert $|e|\sigma_{\rm SHE}$ to the three-dimensional spin Hall conductivity, we divide
$|e|\sigma_{\rm SHE}$ by the interlayer distance,\cite{Inokuchi,Kitou} $\ell=1.77$nm. 
As a result, we obtain
\begin{equation}
|e|\sigma^{\rm 3D}_{\rm SHE} = \frac{|e|\sigma_{\rm SHE}}{\ell} 
\sim 220 \ \Omega^{-1} {\rm cm}^{-1}, 
\end{equation}
which is comparable to Pt (240 $\Omega^{-1} {\rm cm}^{-1}$).

In the case without tilting ($t=0$) and at $T=0$, we find an analytic expression 
for any value of $\Gamma$ as\cite{SM}
\begin{equation}\begin{split}
\sigma_{\rm SHE} 
&= \frac{|e|}{2\pi^2 \hbar} \biggl[
\left( 1+ \frac{\Delta}{\mu} \right) \tan^{-1}\frac{\mu+\Delta}{\Gamma} 
- \left(1-\frac{\Delta}{\mu} \right)  \tan^{-1} \frac{\mu-\Delta}{\Gamma}\biggr].
\end{split}\end{equation}
As shown in the inset of Fig.~\ref{f1}, the effect of finite $\Gamma$ is smearing of the 
chemical potential dependences.

Next, we consider the linear order with respect to the applied magnetic field.
On the basis of Fukuyama's formula for the gauge-invariant Hall conductivity,\cite{FukuyamaHE} 
recently, we have extended such calculations to $\sigma_{\mu\nu}^{(1)}$ 
with the help of $1/\Gamma$ expansion:\cite{Konye}
\begin{equation}\begin{split}
\sigma_{xy}^{(1)} &= \frac{1}{L^2} \sum_{{\bm k},\sigma_z,\pm} \frac{|e|^3 B}{\hbar^2}
\biggl[ \frac{f'(E_\pm)}{8\Gamma^2} (
\partial_x E_\pm \partial_x E_\pm \partial_y^2 E_\pm \cr
&+\partial_y E_\pm \partial_y E_\pm \partial_x^2 E_\pm
-2\partial_x E_\pm \partial_y E_\pm \partial_x \partial_y E_\pm ) \cr
&+\frac{f'(E_\pm)}{4\Gamma} (2\partial_x E_\pm \partial_y E_\pm \Omega_{xy}^\pm
+2 (\partial_x \partial_y E_\pm) M_{xy}^\pm \cr
&-\partial_x E_\pm \partial_y M_{xy}^\pm - \partial_y E_\pm \partial_x M_{xy}^\pm )
\biggr] + O(\Gamma^0), \cr
\sigma_{xx}^{(1)} &= \frac{1}{L^2} \sum_{{\bm k}, \sigma_z,\pm} \frac{|e|^3 B}{\hbar^2}
\frac{f'(E_\pm)}{2\Gamma}( \partial_x E_\pm \partial_x E_\pm \Omega_{xy}^\pm \cr
&+\partial_x^2 E_\pm M_{xy}^\pm -\partial_x E_\pm \partial_x M_{xy} )+ O(\Gamma^0), \cr
\sigma_{yy}^{(1)} &= \frac{1}{L^2} \sum_{{\bm k}, \sigma_z, \pm} \frac{|e|^3 B}{\hbar^2}
\frac{f'(E_\pm)}{2\Gamma} ( \partial_y E_\pm \partial_y E_\pm \Omega_{xy}^\pm \cr
&+\partial_y^2 E_\pm M_{xy}^\pm -\partial_y E_\pm \partial_y M_{xy} )+ O(\Gamma^0),
\end{split}\end{equation}
where the orbital magnetic moment $M_{xy}^\pm$ is given by\cite{Shi,Konye} 
\begin{equation}
M_{xy}^\pm =\frac{1}{2} \frac{{\bm h}\cdot (\partial_x {\bm h}
\times \partial_y {\bm h})}{h^2} = v^2 \hbar^2 \frac{\sigma_z \Delta}{2E_{\bm k}^2}. 
\end{equation}
First, we discuss the conventional Hall conductivity $\sigma_{xy}^{(1)}$. 
Substituting $\Omega_{xy}^\pm$ and $M_{xy}^\pm$,
we can see that the contributions in $\sigma_{xy}^{(1)}$ related to 
$\Omega_{xy}^\pm$ and $M_{xy}^\pm$ exactly vanish.
This means that the contributions from the topological properties 
(i.e., $\Omega_{xy}^\pm$ and $M_{xy}^\pm$) do not appear in $\sigma_{xy}^{(1)}$, 
and that $\sigma_{xy}^{(1)}$ can be understood as the Drude-type Hall conductivity. 
%
In contrast, $\sigma_{xx}^{(1)}$ and $\sigma_{yy}^{(1)}$ are written solely in terms of 
$\Omega_{xy}^\pm$ and $M_{xy}^\pm$. 
However, they are exactly zero due to the spin summation, as in $\sigma_{xy}^{(0)}$.

\begin{figure}
\includegraphics[scale=0.4]{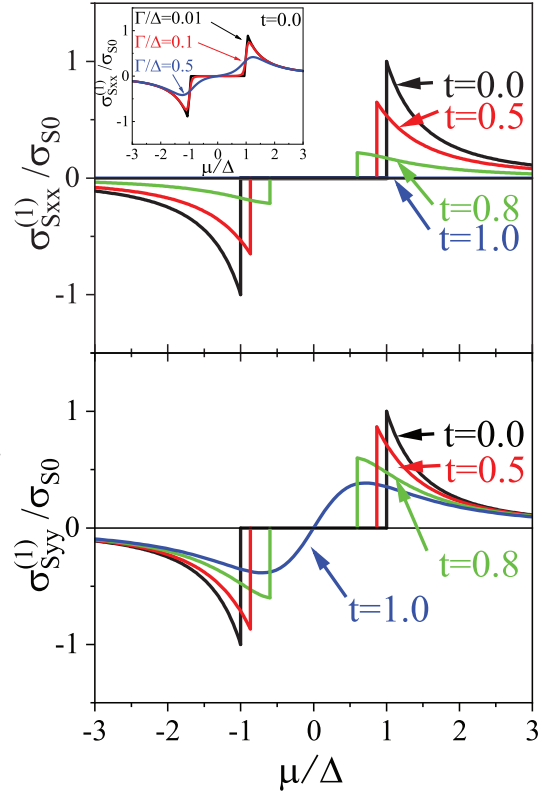} 
\vskip -0.5truecm
\caption{(Color online) Chemical potential dependences of the diagonal spin magnetoconductivity 
$\sigma_{Sxx}^{(1)}$ and $\sigma_{Syy}^{(1)}$ per valley at $T=0$ 
for some values of the tilting parameter $t$ ($t=0, 0.5, 0.8$ and $1.0$). 
The unit of the $y$-axis is $\sigma_{\rm S0}=e^2Bv^2/4\pi \Gamma\Delta$.
(Inset) $\Gamma$-dependence for the case $t=0$.}
\label{f2}
\end{figure}

Again, when we consider the spin magnetoconductivity, we obtain the 
finite values of $\sigma_{Sxx}^{(1)}$ and $\sigma_{Syy}^{(1)}$.
They are anomalous magnetoconductivities in the sense that they are 
odd functions of the applied magnetic field $B$ even if they are diagonal conductivity.
In the present case, using Eq.~(\ref{BerryM}), we obtain
\begin{equation}\begin{split}
\sigma_{Sxx}^{(1)} &= (1-t^2) \sigma_{Syy}^{(1)}, \cr
\sigma_{Syy}^{(1)} &= -\frac{e^2Bv^4\hbar^2 \Delta}{4\Gamma L^2} 
\sum_{{\bm k},\sigma_z, \pm} (\pm)
\frac{f'(E_\pm)}{E_{\bm k}^3} +O(\Gamma^0).
\label{eq:Sxxyy}
\end{split}\end{equation}
At $T=0$, the summation over $\bm k$ can be carried out, which leads to\cite{SM}
\begin{equation}
\sigma_{Syy}^{(1)} = \frac{e^2 B v^2}{4\pi \Gamma}  
\frac{\mu \Delta}{(\mu^2 + t^2 \Delta^2)^{3/2}}
\theta(|\mu|-\sqrt{1-t^2}\Delta) +O(\Gamma^0).
\label{eq:Syy1}
\end{equation}
Figure \ref{f2} shows the chemical potential dependences of 
$\sigma_{Sxx}^{(1)}$ and $\sigma_{Syy}^{(1)}$ for some values of the tilting parameter $t$. 
Note that $\sigma_{Sxx}^{(1)}$ and $\sigma_{Syy}^{(1)}$ vanish when the chemical potential is inside the gap that shrinks as 
$t$ increases. 
$\sigma_{Sxx}^{(1)}$ becomes smaller with increasing $t$, but $\sigma_{Syy}^{(1)}$ continues 
to be of the same order as in the case with $t=0$. 
For $t=0$ and at $T=0$, we find an analytic expression, which is shown in the 
Supplemental Material.\cite{SM}
Again, the effect of finite $\Gamma$ is smearing of the chemical potential dependences,
as shown in the inset of Fig.~\ref{f2}.

When the chemical potential $\mu$ is located at the bottom of the upper Dirac cone, 
i.e., $\mu=\sqrt{1-t^2}\Delta$, $\sigma_{Sxx}^{(1)}$ and $\sigma_{Syy}^{(1)}$ become
\begin{equation}
\sigma_{Sxx}^{(1)} = (1-t^2) \sigma_{Syy}^{(1)}, \quad
\sigma_{Syy}^{(1)} = \frac{e^2 B v^2}{4\pi \Gamma\Delta} \sqrt{1-t^2}.
\label{SxxFinal}
\end{equation}
Let us estimate these values. 
As typical parameters, we use $v\sim 1.2 \times 10^5$ m/s, $t \sim 0.8$, and the energy gap $=2$ meV, 
obtained in organic conductors.\cite{Kitou,TsumuSuzu,OrganicRev}
Because the gap in the present model is given by $2\Delta\sqrt{1-t^2}$, we have 
$\Delta=1.7$ meV. 
Assuming $\Gamma=3$ K $\sim 0.3$ meV as a typical value of the relaxation 
rate,\cite{OrganicRev,SuzuOga} we obtain
\begin{equation}
\frac{|e|\sigma_{Sxx}^{(1)}}{\ell} \sim 430 B\ \Omega^{-1} {\rm cm}^{-1}, \
\frac{|e|\sigma_{Syy}^{(1)}}{\ell} \sim 1200 B\ \Omega^{-1} {\rm cm}^{-1},
\label{SxxEval}
\end{equation}
at $\mu=\sqrt{1-t^2}\Delta$, where $B$ is measured in Tesla.
[Note that the Zeeman interaction gives a similar spin magnetoconductivity. 
However, its 
magnitude is smaller than Eq.~(\ref{eq:Sxxyy}) by a factor $\Delta/mv^2\sim 0.013$ 
for the present parameters.\cite{SM}]

\begin{figure}
\includegraphics[scale=0.35]{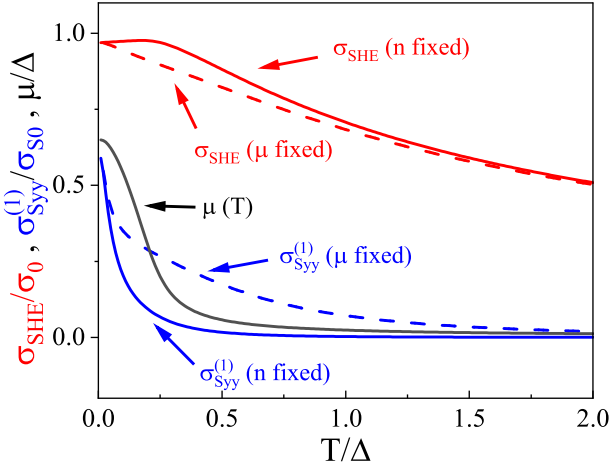} 
\caption{(Color online) Temperature dependences of chemical potential $\mu$, $\sigma_{\rm SHE}$, 
and $\sigma_{Syy}^{(1)}$ per valley for $t=0.8$ and a fixed value of $n$. 
For comparison, the case of a fixed chemical potential ($\mu/\Delta$ is fixed at 0.65) 
is also shown by dashed lines.}
\label{f3}
\end{figure}

Figure \ref{f3} shows the temperature dependences of $\mu$, 
$\sigma_{\rm SHE}$, and $\sigma_{Syy}^{(1)}$ for $t=0.8$ and a fixed value of the 
electron density $n$, where the zero-temperature chemical potential $\mu_0$ 
is assumed to be $\mu_0/\Delta=0.65$. 
With increasing temperature, 
$\sigma_{\rm SHE}$ only slightly decreases, but $\sigma_{Syy}^{(1)}$ 
(and $\sigma_{Sxx}^{(1)}$) decreases rather rapidly because the chemical potential 
approaches 0 and the contribution of the holes in the valence band tends to cancel that of 
electrons. This is a characteristic feature in the perfectly symmetric bands assumed here.
However, when there is asymmetry in the density of states, the temperature 
dependence of $\mu$ can be different from the case of 
constant $n$.\cite{KobayashiFuku,SuzuTsumu} 
As an example of such a case, we calculate the case of a fixed chemical potential 
($\mu/\Delta$ is fixed at 0.65), which is shown by dashed lines in Fig.~\ref{f3}.
In this case, $\sigma_{Syy}^{(1)}$ does not decrease rapidly, 
whereas the behavior of $\sigma_{\rm SHE}$ changes only slightly. 
Note that the resistivity increases below 50 K in $\alpha$-(BETS)$_2$I$_3$, 
whose origin is not understood well.\cite{OhkiKoba}
Therefore, to apply the present result to actual materials, we must take into account 
the low-temperature electronic states. 


In summary, we studied anomalous spin transport properties in a gapped Dirac electron system.
The gap opening due to the spin--orbit interaction leads to a spin Hall effect and 
diagonal magnetoconductivity proportional to the applied magnetic field $B$, which 
will be observed experimentally by using the inverse spin Hall effect. 
The magnitudes of these conductivities are estimated.


We thank K.\ Kitayama, V.\ K\"onye, 
H.\ Maebashi, and S.\ Fujiyama for fruitful discussions. 
This work is supported by Grants-in-Aid for Scientific Research from 
JST-Mirai Program Grant Number JPMJMI19A1, Japan, 
and the Japan Society for the
Promotion of Science (No.\ JP18H01162, 19K03720, 18K03482). 

\def\journal#1#2#3#4{#1 {\bf #2}, #3 (#4)}
\def\PR{Phys.\ Rev.}
\def\PRB{Phys.\ Rev.\ B}
\def\PRL{Phys.\ Rev.\ Lett.}
\def\JPSJ{J.\ Phys.\ Soc.\ Jpn.}
\def\PTP{Prog.\ Theor.\ Phys.}
\def\JPCS{J.\ Phys.\ Chem.\ Solids}
\def\PRR{Phys.\ Rev.\ Research}

\end{document}


\maketitle

\vskip-2truecm

\noindent {\bf 1. Derivation of Eq.~(5) in the main text}

In general, the conductivity tensor is obtained by
\begin{equation}
\sigma_{\mu\nu}= \lim_{\omega\rightarrow 0} \frac{1}{i(\omega+i\delta)} 
\left\{ \tilde\Phi_{\mu\nu} (\omega) - \tilde\Phi_{\mu\nu} (0) \right\}, 
\label{eq:Conduc}
\end{equation}
for $\mu, \nu = x,y,z$, where $\tilde\Phi_{\mu\nu} (\omega)$ is the analytic continuation of 
\begin{equation}
\Phi_{\mu\nu} (i\omega_\lambda) = \frac{1}{V} \int_0^\beta d\tau 
\langle j_{\mu} (\tau) j_{\nu} (0)\rangle e^{i\omega_\lambda \tau}, 
\label{ThermalCorr}
\end{equation}
with $i\omega_\lambda \rightarrow \hbar(\omega + i\delta)$. 
Here, $V$ is the volume of the system in general 
($V=L^2$ in two dimensions for the case of the main text), 
$\beta=1/k_{\rm B}T$, $\omega_\lambda = 2\pi \lambda k_{\rm B}T$ is the Matsubara
frequency for the external field with $\lambda$ being an integer, and $j_{\mu}(\tau)$ 
is the $\mu$-component of the electric current. 
With the periodic part of the Bloch wave function $\Uell({\bm r})$ that satisfies 
\begin{equation}
H_{\bm k} \Uell({\bm r}) = \Eell({\bm k}) \Uell({\bm r}),
\label{UellEq}
\end{equation}
with $H_{\bm k} = e^{-i{\bm k}\cdot {\bm r}} H e^{i {\bm k}\cdot {\bm r}}$, 
the matrix element of the current operator between the band indices $a$ and $b$ 
becomes\cite{Wilson,Ogata}
\begin{equation}
({j}_{\mu})_{ab}  
= \frac{e}{\hbar} \int u^\dagger_{a,{\bm k}} \Hkm u^{\phantom{\dagger}}_{b,{\bm k}} d{\bm r} 
= \frac{e}{\hbar} \langle a \bigl| \Hkm \bigr| b \rangle 
\equiv \frac{e}{\hbar} (\gamma_\mu)_{ab},
\label{eq:JmatrixEl}
\end{equation}
where $\langle a |$ and $|b\rangle $ are the abbreviations of $u^\dagger_{a,{\bm k}}$ 
and $u_{b,{\bm k}}$, respectively. 
Using the $k_\mu$-derivative of eq.~(\ref{UellEq}), 
\begin{equation}
\left( \frac{\partial \varepsilon_{a}}{\partial k_\mu} - \frac{\partial H_{\bm k}}{\partial k_\mu} 
\right)  |a\rangle + (\varepsilon_a - H_{\bm k}) |\partial_\mu a \rangle =0,
\end{equation}
with
$| \partial_\mu a \rangle = \frac{\partial}{\partial k_\mu} u_{a,{\bm k}}$,
we obtain
\begin{equation}
(\gamma_\mu)_{ab} = \frac{\partial \varepsilon_{a}}{\partial k_\mu}\delta_{ab}
+ (\varepsilon_b - \varepsilon_a )\langle a | \partial_\mu b \rangle.
\label{eq:gamma_matrix_element}
\end{equation}

The current--current correlation function (\ref{ThermalCorr}) becomes
\begin{equation}
\Phi_{\mu\nu} (i\omega_\lambda) =-\frac{k_{\rm B}T}{V} \sum_{n,{\bm k}} 
\frac{e^2}{\hbar^2} {\rm Tr} \left[ \gamma_{\mu} {\mathcal G}_+ \gamma_\nu {\mathcal G}\right],
\label{PhiJJ0}
\end{equation}
where vertex corrections have been neglected, and 
the trace (Tr) is taken over the band index including spin.
The thermal Green's function is band-index diagonal and has the form 
\begin{equation}
{\mathcal G}_a ({\bm k}, \varepsilon_n) = \frac{1}{i\varepsilon_n - \Eell({\bm k}) + \mu 
+i\Gamma {\rm sign} (\varepsilon_n)},
\end{equation}
where $\varepsilon_n=(2n+1)\pi k_{\rm B}T$ is the fermion Matsubara frequency, 
and the relaxation rate $\Gamma$ has been used by assuming the simple impurity scattering. 
In (\ref{PhiJJ0}), we have used abbreviations 
${\mathcal G}({\bm k}, i\varepsilon_n) \rightarrow {\mathcal G}$ and 
${\mathcal G}({\bm k}, i\varepsilon_n+i\omega_\lambda) \rightarrow {\mathcal G}_+$. 
With the matrix element (\ref{eq:gamma_matrix_element}), Eq.~(\ref{PhiJJ0}) becomes
\begin{equation}
\Phi_{\mu\nu}(i\omega_\lambda) 
=-\frac{k_{\rm B}T}{V} \sum_{n, {\bm k}} 
\frac{e^2}{\hbar^2} \biggl[ \sum_a 
\frac{\partial \varepsilon_{a}}{\partial k_\mu}\frac{\partial \varepsilon_{a}}{\partial k_\nu}
{\mathcal G}_{a,+} {\mathcal G}_a 
- \sum_{a,b} (\varepsilon_a-\varepsilon_b)^2 \langle a|\partial_\mu b\rangle \langle b |\partial_\nu a \rangle 
{\mathcal G}_{b,+} {\mathcal G}_a \biggr],
\end{equation}
where the summation over $a$ contains spin.
The first (second) term represents the intra- (inter-) band contribution. 
The summation over the Matsubara frequency can be taken in a standard way, and then 
the linear-$\omega$ term as in (\ref{eq:Conduc}) is extracted. 
As a result, the conductivity becomes
\begin{equation}
\sigma_{\mu\nu}  =-\frac{1}{V} \sum_{{\bm k}} \frac{2e^2}{\hbar} {\rm Re} \biggl[ \sum_a 
\frac{\partial \varepsilon_{a}}{\partial k_\mu}\frac{\partial \varepsilon_{a}}{\partial k_\nu}
C_{aa} 
- \sum_{a,b} (\varepsilon_a-\varepsilon_b)^2 \langle a|\partial_\mu b\rangle \langle b |\partial_\nu a \rangle 
C_{ba} \biggr],
\label{eq:sigma0}
\end{equation}
with
\begin{equation}
C_{ba} := \int \frac{d\varepsilon}{2\pi} f(\varepsilon) \frac{\partial 
G_b^{\rm R}}{\partial \varepsilon} (G_a^{\rm R}-G_a^{\rm A}), 
\label{eq:Cab}
\end{equation}
where $f(\varepsilon) = \frac{1}{e^{\beta(\varepsilon-\mu)}+1}$, 
the retarded Green's function $G_a^{{\rm R}}$ 
is defined by 
$G_a^{{\rm R}} = \frac{1}{{\varepsilon-\varepsilon_a ({\bm k}) +i\Gamma }}$, and 
the advanced Green's function is $G_a^{{\rm A}}= [G_a^{{\rm R}}]^*$.
Then, we assume that the relaxation rate $\Gamma$ is sufficiently small to make 
the $\Gamma^\ell$ expansion with $\ell$ being an integer, 
and we obtain\cite{Konye}
\begin{equation}
C_{ba} = \begin{cases}
- \frac{i f(\varepsilon_a)}{4 \Gamma^2} + \frac{f'(\varepsilon_a)}{4\Gamma}
+ \frac{i f''(\varepsilon_a)}{8} + O(\Gamma^1), &{\rm for}\ a=b, \cr
\frac{i f(\varepsilon_a)}{(\varepsilon_a-\varepsilon_b)^2} + O(\Gamma^1), &{\rm for}\ a\ne b. \end{cases}
\end{equation}
With this small $\Gamma$ expansion, Eq.~(\ref{eq:sigma0}) becomes
\begin{equation}
\sigma_{\mu\nu}
=-\frac{1}{V} \sum_{{\bm k}} \frac{2e^2}{\hbar} 
\biggl[ \sum_a \frac{f'(\varepsilon_a)}{4\Gamma} 
\frac{\partial \varepsilon_{a}}{\partial k_\mu}\frac{\partial \varepsilon_{a}}{\partial k_\nu}
-\sum_{a,b} f(\varepsilon_a) {\rm Im} \left\{ \langle \partial_\mu a| b\rangle \langle b |\partial_\nu a \rangle \right\} 
+ O(\Gamma^1) \biggr],
\label{eq:sigma0B}
\end{equation}
where we have used the relation 
$\langle a| \partial_\mu b \rangle = - \langle \partial_\mu a |b \rangle$. 
The first term gives the Drude-type conductivity. 
The summation over $b$ in the second term
can be taken by using the completeness condition, which leads to
\begin{equation}
\frac{1}{V} \sum_{a,{\bm k}} \frac{2e^2}{\hbar} f(\varepsilon_a)
{\rm Im} \langle \partial_\mu a |\partial_\nu a \rangle
= -\frac{1}{V} \sum_{a,{\bm k}} \frac{e^2}{\hbar} f(\varepsilon_a) \Omega_{a, \mu\nu},
\end{equation}
where $\Omega_{a, \mu\nu}$ is the Berry curvature for the band $a$,
$\Omega_{a, \mu\nu} = i \{ \langle \partial_\mu a |\partial_\nu a \rangle 
- \langle \partial_\nu a |\partial_\mu a \rangle \}.
$
This is the contribution obtained by Karplus--Luttinger\cite{KarplusLuttinger}
to discuss the anomalous Hall conductivity although the term 
\lq\lq Berry curvature'' was not known at that time.
Alternatively, if we use the matrix element in (\ref{eq:JmatrixEl}), 
the second term can be rewritten as 
\begin{equation}
\frac{1}{V} \sum_{a\ne b, {\bm k}} \frac{2e^2}{\hbar}f(\varepsilon_a)  {\rm Im}
\frac{\langle a|\partial_\mu H |b\rangle \langle b |\partial_\nu H|a \rangle}{(\varepsilon_a-\varepsilon_b)^2},
\label{eq:sigmaHall}
\end{equation}
which is equivalent to the TKNN formula for the quantum Hall effect.\cite{TKNN}
In the model considered in the main text, we have two bands with $\varepsilon_a = E_\pm$. 
In this case, we obtain Eq.~(5) in the main text. 

\noindent {\bf 2. Momentum integrals of $\sigma_{\rm SHE}$ and $\sigma_{Syy}^{(1)}$}

We will show the calculation for the case of $\mu<-\Delta\sqrt{1-t^2}$. At $T=0$, we have
\begin{equation}
\sigma_{\rm SHE} = 
\sigma_{Sxy}^{(0)}=\frac{|e|\Delta}{8\pi^2\hbar} \sum_{\sigma_z} \iint dK_x dK_y \frac{1}{E_{\bm k}^3} 
\theta( \mu + tK_x+E_{\bm k}), \label{SM:Sxy1}
\end{equation}
where $E_{\bm k} = \sqrt{\Delta^2 + K_x^2+ K_y^2}$ and 
we have put $K_x=v\hbar k_x$ and $K_y = v\hbar k_y$.
The $K_y$-integral yields
\begin{equation}\begin{split}
\sigma_{\rm SHE} &=\frac{|e|\Delta}{8\pi^2\hbar} \sum_{\sigma_z} \int_{K_1}^{K_2} dK_x 
\frac{1}{A_{K_x}} \biggl[ \frac{2}{|\mu+ tK_x|} - \frac{1}{|\mu+ tK_x|+A_{K_x}}
- \frac{1}{|\mu+ tK_x|-A_{K_x}} \biggr] \cr 
&+\frac{|e|\Delta}{8\pi^2\hbar} \sum_{\sigma_z} 
\int_{-\infty}^{\infty} dK_x \frac{2}{\Delta^2+K_x^2}, 
\label{SM:Sxy2}
\end{split}\end{equation}
with 
\begin{equation}\begin{split}
A_{K_x} &= \sqrt{(\mu+tK_x)^2-\Delta^2-K_x^2}, \cr 
K_1 &= \frac{t\mu}{1-t^2} - \frac{r_0}{\sqrt{1-t^2}}, \qquad 
K_2 = \frac{t\mu}{1-t^2} + \frac{r_0}{\sqrt{1-t^2}},  \qquad
r_0 = \sqrt{\frac{\mu^2}{1-t^2}-\Delta^2},
\end{split}\end{equation}
where we have used the relation $K_x^2+\Delta^2=(\mu+tK_x)^2- A_{K_x}^2$. 
The integrals in the first term can be carried out by using the change of 
the variable from $K_x$ to $\theta$ as
\begin{equation}
\sin \theta = \frac{\sqrt{1-t^2}}{r_0} \left( K_x - \frac{t\mu}{1-t^2} \right),
\label{SM:ChangeofV}
\end{equation}
and using the integral, 
$\int_{-\pi/2}^{\pi/2} d\theta/(a\cos\theta+b) = \pi{\rm sign} (b)/\sqrt{b^2-a^2}$
for $|b|>|a|$. 
Then, we obtain Eq.~(7) in the main text. 
Note that in the case of $|\mu|<\Delta\sqrt{1-t^2}$, only 
the last term in (\ref{SM:Sxy2}) exists.

$\sigma_{Syy}^{(1)}$ can be calculated similarly. At $T=0$, we have
\begin{equation}
\sigma_{Syy}^{(1)} = \frac{e^2Bv^2\Delta}{16\pi^2 \Gamma } 
\sum_{\sigma_z,\pm} (\pm) \iint dK_x dK_y 
\frac{\delta(-tK_x\pm E_{\bm k}-\mu)}{E_{\bm k}^3}.
\end{equation}
The $K_y$-integral leads to 
\begin{equation}
\sigma_{Syy}^{(1)} = \frac{e^2Bv^2\Delta}{8\pi^2 \Gamma } 
\sum_{\sigma_z} \int
\frac{dK_x}{A_{K_x} |\mu + tK_x|^2} {\rm sign} (\mu) \theta(|\mu|-\sqrt{1-t^2}\Delta). 
\end{equation}
Again, the $K_x$ integral can be carried out by using (\ref{SM:ChangeofV}) 
and $\int_{-\pi/2}^{\pi/2} d\theta/(a\cos\theta+b)^2 = \pi |b|/({b^2-a^2})^{3/2}$, which gives
Eq.~(13) in the main text. 

\noindent {\bf 3. Finite $\Gamma$ calculation}

For the case of finite $\Gamma$, it is convenient to start from the matrix form 
of $\Phi_{\mu\nu} (i\omega_\lambda)$ in (\ref{PhiJJ0}) and 
to use the explicit form of the thermal Green's function in the present Hamiltonian:
\begin{equation}
\label{eq:ThermalG}
\mathcal{G}({\bm k}, i\varepsilon_n) =\frac{i\varepsilon_n + tK_x +\mu
+ K_x {\tau}_x +K_y \tau_y + \sigma_z \Delta \tau_z}{(i\varepsilon_n + tK_x+\mu)^2-E_{\bm k}^2}.
\end{equation}
Again, we have put $K_x=v\hbar k_x$ and $K_y = v\hbar k_y$. 
In the following, we consider the $t=0$ case. Using the matrix form of current, 
$j_{Sx} = v\sigma_z \tau_x$ and $j_y = ev\tau_y$, taking the trace (Tr), and taking the linear order of 
$i\omega_\lambda$, we obtain
\begin{equation}
\label{eq:KuboXX}
\sigma_{\rm SHE} =\frac{2i|e|v^2\hbar \Delta}{L^2} \sum_{{\bm k}, \sigma_z} 
\int_{-\infty}^\infty \frac{d\varepsilon}{2\pi} f(\varepsilon) \biggl[ 
\frac{1}{D_R^2}-2i\Gamma \frac{\partial}{\partial \varepsilon} \left( \frac{1}{D_R D_A} \right) 
-\frac{1}{D_A^2} \biggr],
\end{equation}
with $D_R = (\varepsilon+i\Gamma)^2-E_{\bm k}^2$ and 
$D_A = (\varepsilon-i\Gamma)^2-E_{\bm k}^2$. 
The $K_x$ and $K_y$ integrals lead to
\begin{equation}\begin{split}
\sigma_{\rm SHE} =\frac{i|e|\Delta}{2\pi\hbar} \sum_{\sigma_z} 
\int_{-\infty}^\infty \frac{d\varepsilon}{2\pi} f(\varepsilon) &\biggl[ 
-\frac{1}{(\varepsilon+i\Gamma)^2-\Delta^2}+\frac{1}{(\varepsilon-i\Gamma)^2-\Delta^2} \cr
&- i \frac{\partial}{\partial \varepsilon} \left\{ \frac{1}{\varepsilon} 
\left( \tan^{-1} \frac{\varepsilon+\Delta}{\Gamma} + \tan^{-1} \frac{\varepsilon-\Delta}{\Gamma} 
 \right) \right\} \biggr].
\end{split}\end{equation}
At $T=0$, the $\varepsilon$-integral can be carried out, which yields
\begin{equation}\begin{split}
\sigma_{\rm SHE} &=\frac{i|e|\Delta}{4\pi^2\hbar} \sum_{\sigma_z} 
\biggl[ -\frac{1}{2\Delta} \ln \frac{(\mu+i\Gamma-\Delta)(\mu-i\Gamma+\Delta)}
{(\mu+i\Gamma+\Delta)(\mu-i\Gamma-\Delta)}
- \frac{i}{\mu} 
\left( \tan^{-1} \frac{\mu+\Delta}{\Gamma} + \tan^{-1} \frac{\mu-\Delta}{\Gamma} 
 \right) \biggr] \cr
&=\frac{|e|}{4\pi^2\hbar} \sum_{\sigma_z} 
\biggl[ \left( 1+\frac{\Delta}{\mu} \right) \tan^{-1} \frac{\mu+\Delta}{\Gamma} 
-\left( 1-\frac{\Delta}{\mu} \right) \tan^{-1} \frac{\mu-\Delta}{\Gamma} \biggr].
\end{split}\end{equation}

In the first order of the magnetic field, we can use Fukuyama's formula\cite{FukuyamaHE}
and its extension.\cite{Konye}
Using the matrix form of the thermal Green's function, we obtain
\begin{equation}
\sigma_{Sxx}^{(1)} =\sigma_{Syy}^{(1)}= \frac{4ie^2 B v^4 \hbar^2 \Gamma \Delta}{L^2} \sum_{{\bm k}, \sigma_z} 
\int_{-\infty}^\infty \frac{d\varepsilon}{2\pi} f'(\varepsilon) \biggl[ 
\frac{1}{D_R D_A^2}-\frac{1}{D_R^2 D_A} \biggr].
\end{equation}
At $T=0$, the $K_x$ and $K_y$ integrals lead to
\begin{equation}
\sigma_{Sxx}^{(1)} = \sigma_{Syy}^{(1)} = \frac{e^2 B v^2 \Delta}{4\pi^2 \mu^2\Gamma} 
\biggl[ \tan^{-1}\frac{\mu+\Delta}{\Gamma}+\tan^{-1} \frac{\mu-\Delta}{\Gamma} 
+2{\rm Re} \frac{\mu\Gamma}{(\mu+i\Gamma)^2-\Delta^2} \biggr].
\label{SxxAnalytic}
\end{equation}
If we make the $1/\Gamma$ expansion of (\ref{SxxAnalytic}), we obtain
\begin{equation}
\sigma_{Sxx}^{(1)} = \sigma_{Syy}^{(1)} = \frac{e^2 B v^2 \Delta}{4\pi \mu^2\Gamma}
\theta (|\mu|-\Delta) {\rm sign}\mu +O(\Gamma^1),
\end{equation}
which coincides with Eq.~(13) in the main text at $t=0$. 


\noindent {\bf 4. Spin magnetoconductivity due to Zeeman interaction}

To estimate the order of magnitude, we consider the case with $t=0$. 
The conventional diagonal conductivity is given by
\begin{equation}
\sigma_{xx} = -\frac{e^2}{\hbar L^2} \sum_{\bm k, \sigma_z, \pm} 
\frac{f'(\pm E_{\bm k})}{2\Gamma} \left( \frac{\partial E_{\bm k}}{\partial k_x} \right)^2.
\end{equation}
When we take into account the Zeeman interaction, we can find its contribution 
to the spin conductivity as
\begin{equation}\begin{split}
\sigma_{Sxx}^{\rm Zeeman} &= \frac{|e|}{\hbar L^2} \sum_{\bm k, \sigma_z, \pm} 
\sigma_z \frac{f'(\pm E_{\bm k}+\frac{|e|\hbar \sigma_z B}{2m})}{2\Gamma} 
\left( \frac{\partial E_{\bm k}}{\partial k_x} \right)^2 \cr
&= \frac{e^2 B}{2m L^2} \sum_{\bm k, \sigma_z, \pm} 
\frac{f''(\pm E_{\bm k})}{2\Gamma}  \left( \frac{\partial E_{\bm k}}{\partial k_x} \right)^2
+O(B^2).
\end{split}\end{equation}
At $T=0$, $f''(\pm E_{\bm k})= -\delta'(\pm E_{\bm k}-\mu)$, and we obtain
\begin{equation}
\sigma_{Sxx}^{\rm Zeeman} 
= \frac{e^2 B}{8\pi m \Gamma} \left( 1+\frac{\Delta^2}{\mu^2} \right) \theta(|\mu|-\Delta) 
+O(B^2).
\end{equation}
At $\mu\sim \Delta$, 
this is smaller by a factor of $\Delta/mv^2$ than the Berry curvature contribution 
of Eq.~(13) in the main text. 
The effective mass in the gapped Dirac electron is $m^*=\Delta/v^2=0.013m$ in the 
present parameters ($\Delta=1.7$ meV and $v=1.2 \times 10^{5}$ m/s). 
Thus, the ratio is $\Delta/mv^2=m^*/m=0.013$.

\def\journal#1#2#3#4{#1 {\bf #2}, #3 (#4)}
\def\PR{Phys.\ Rev.}
\def\PRB{Phys.\ Rev.\ B}
\def\PRL{Phys.\ Rev.\ Lett.}
\def\JPSJ{J.\ Phys.\ Soc.\ Jpn.}
\def\PTP{Prog.\ Theor.\ Phys.}
\def\JPCS{J.\ Phys.\ Chem.\ Solids}
\def\PRR{Phys.\ Rev.\ Research}